\documentstyle[12pt,epsf,rotating]{article}

\font\tenrm=cmr10

\begin{document}
\renewcommand{\thefootnote}{\fnsymbol{footnote} }

\begin{flushright}
PM--97--32 \\
October 1997 \\
\end{flushright}

\vspace*{1cm}

\vglue 0.7cm
\begin{center}{{\bf THE SEARCH FOR THE STANDARD MODEL HIGGS BOSON}\\
\vglue 0.5cm
{\bf AT PRESENT AND FUTURE COLLIDERS}\\
\vglue 1.0cm
{\sc Abdelhak Djouadi}\\
\vglue 0.5cm 
{\it Laboratoire de Physique Math\'ematique et Th\'eorique, 
UPRES--A 5032, \\ 
Universit\'e de Montpellier II, F--34095 Montpellier Cedex 5, France.}\\

\vglue 1.5cm
{\sc Abstract}}
\end{center}
{\rightskip=3pc
 \leftskip=3pc
\tenrm\baselineskip=14pt

\noindent I briefly review the Higgs sector in the Standard Model.
After summarizing the properties of the Higgs boson, I will discuss 
the prospects for discovering this particle at the present colliders 
LEP2 and Tevatron and at the next generation colliders LHC and a 
high--energy $e^+e^-$ linear collider. The possibilities of studying 
the properties of the Higgs particle will be then summarized. 

\vspace*{3cm}

\normalsize

\noindent Contribution to the European Network Workshop {\it Tests of 
the Electroweak Symmetry Breaking}, Ouranoupolis, Greece, May 1997. 
}

\newpage

\newcommand{\s}{\\ \vspace*{-3mm} }
\newcommand{\nn}{\noindent}
\newcommand{\non}{\nonumber}
\newcommand{\ee}{e^+ e^-}
\newcommand{\ra}{\rightarrow}
\newcommand{\lra}{\longrightarrow}
\newcommand{\beq}{\begin{eqnarray}}
\newcommand{\eeq}{\end{eqnarray}}
\newcommand{\tb}{{\rm tg} \beta}

\newcommand{\lsim}{\raisebox{-0.13cm}{~\shortstack{$<$ \\[-0.07cm] $\sim$}}~}
\newcommand{\gsim}{\raisebox{-0.13cm}{~\shortstack{$>$ \\[-0.07cm] $\sim$}}~}

\setcounter{page}{2}
\clearpage
\def\thefootnote{\arabic{footnote}}
\setcounter{footnote}{0}

\subsection*{1. The Standard Higgs Boson}

\subsubsection*{1.1 Introduction} 

The search for Higgs particles is one of the main missions of present
and future high--energy colliders. The observation of this particle is
of utmost importance for the present understanding of the interactions
of the fundamental particles. Indeed, in order to accomodate the
well--established electromagnetic and weak interaction phenomena, the
existence of at least one isodoublet scalar field to generate fermion
and weak gauge bosons masses is required. The Standard Model (SM) makes
use of one isodoublet field: three Goldstone bosons among the four 
degrees of freedom are absorbed to build up the longitudinal components of 
the massive $W^\pm,Z$ gauge bosons; one degree of freedom is left over 
corresponding to a physical scalar particle, the Higgs boson \cite{R1}. 
Despite of its numerous successes in explaining the present data, the 
Standard Model will not be completely tested before this particle has 
been experimentally observed and its fundamental properties studied. \s

In the Standard Model, the mass of the Higgs particle is a free
parameter. The only available information is the upper limit $M_H \gsim 77$
GeV established at LEP2 \cite{R2}, although the high--precision
electroweak data from LEP and SLC seem to indicate that its mass is
smaller than a few hundred GeV \cite{R3}. However, interesting
theoretical constraints can be derived from assumptions on the energy
range within which the model is valid before perturbation theory breaks
down and new phenomena would emerge: \s

-- If the Higgs mass were larger than $\sim$ 1 TeV, the $W$ and $Z$ 
bosons would interact strongly with each other to ensure unitarity 
in their scattering at high energies. \s

-- The quartic Higgs self--coupling, which at the scale $M_H$ is
fixed by $M_H$ itself, grows logarithmically with the energy scale. If
$M_H$ is small, the energy cut--off $\Lambda$ at which the coupling
grows beyond any bound and new phenomena should occur, is large;
conversely, if $M_H$ is large, $\Lambda$ is small. The condition $M_H 
\lsim \Lambda$ sets an upper limit on the Higgs mass in the SM; lattice
analyses lead to an estimate of about 630 GeV for this limit.
Furthermore, top quark loops tend to drive the coupling to negative
values for which the vacuum is no more stable. Therefore, requiring the
SM to be extended to the GUT scale, $\Lambda_{\rm GUT} \sim 10^{15}$
GeV, and including the effect of top quark loops on the running
coupling, the Higgs boson mass should roughly lie in the range between 
100 and 200 GeV; see the account given in Ref.~\cite{R4}. \s

The search for the Higgs particle will be the major goal of the next
generation of colliders. In the following, after summarizing the properties
of the Higgs boson, I will briefly discuss the discovery potential of the 
present colliders LEP2 \cite{R4} and Tevatron \cite{R5} as well as the pp 
collider LHC \cite{R6,X6} with a c.m. energy of $\sim 14$ TeV and a future 
${\rm \ee}$ linear collider \cite{R7,X7} with a c.m. energy in 
the range of 300 to 500 GeV. A more detailed
discussion and a complete set of references can be found in Refs.~[4--7].

\subsubsection*{1.2 Decay Modes} 

In the SM, the profile of the Higgs particle is uniquely determined once
$M_H$ is fixed. The decay width, the branching ratios and the production
cross sections are given by the strength of the Yukawa couplings to
fermions and gauge bosons, the scale of which is set by the masses of
these particles. To discuss the Higgs decay modes \cite{R8}, it is convenient 
to divide the Higgs mass into two ranges: the ``low mass" range $M_H \lsim 130$
GeV and the ``high mass" range $M_H \gsim 130$ GeV. \s

In the ``low mass" range, the Higgs boson decays into a large variety of
channels. The main decay mode is by far the decay into $b\bar{b}$ pairs
with a branching ratio of $\sim 90\%$ followed by the decays into
$c\bar{c}$ and $\tau^+\tau^-$ pairs with a branching ratio of $\sim
5\%$. Also of significance, the top--loop mediated Higgs decay into
gluons, which for $M_H$ around 120 GeV occurs at the level of $\sim
5\%$. The top and $W$--loop mediated $\gamma\gamma$ and $Z \gamma$ decay
modes are very rare the branching ratios being of ${\cal O }(10^{-3})$;
however these decays lead to clear signals and are interesting being
sensitive to new heavy particles. \s

In the ``high mass" range, the Higgs bosons decay into $WW$ and $ZZ$
pairs, with one of the gauge bosons being virtual below the threshold.
Above the $ZZ$ threshold, the Higgs boson decays almost exclusively into
these channels with a branching ratio of 2/3 for $WW$ and 1/3 for $ZZ$.
The opening of the $t\bar{t}$ channel does not alter significantly this
pattern, since for large Higgs masses, the $t\bar{t}$ decay width rises
only linearly with $M_H$ while the decay widths to $W$ and $Z$ bosons
grow with $M_H^3$. \s

In the low mass range, the Higgs boson is very narrow $\Gamma_H<10$ MeV,
but the width becomes rapidly wider for masses larger than 130 GeV,
reaching 1 GeV at the $ZZ$ threshold; the Higgs decay width cannot be
measured directly in the mass range below 250 GeV. For large masses, 
$M_H \gsim 500$ GeV, the Higgs boson becomes obese: its decay width becomes 
comparable to its mass.

\begin{figure}[hbtp]
\begin{center}
\vspace*{-1.0cm}

\hspace*{-1.5cm}
\begin{turn}{-90}%
\epsfxsize=9.cm \epsfbox{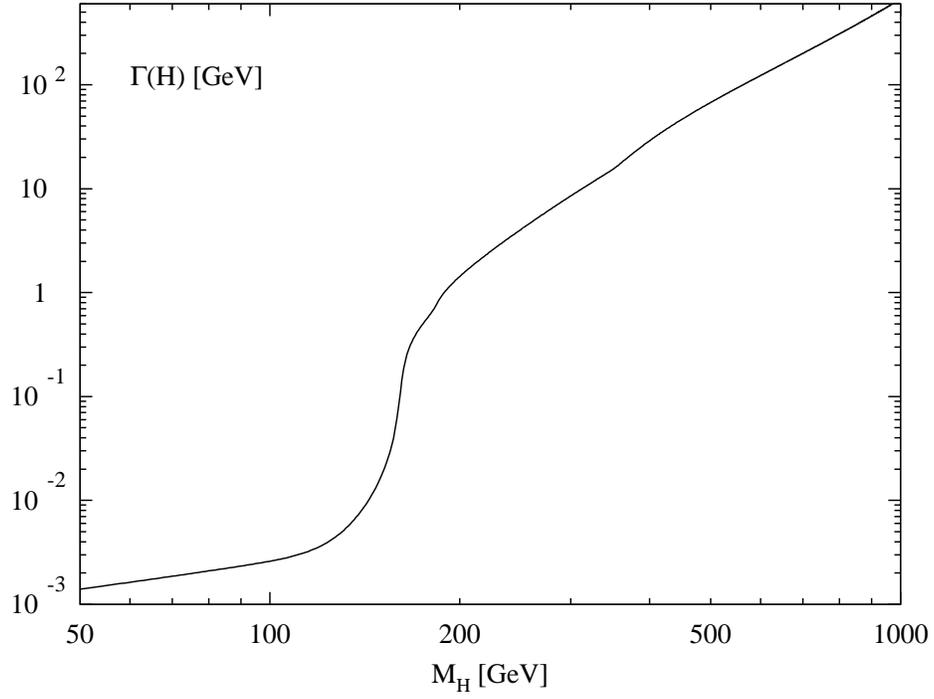}
\end{turn}
\vspace*{1.cm}

\hspace*{-1.5cm}
\begin{turn}{-90}%
\epsfxsize=9.cm \epsfbox{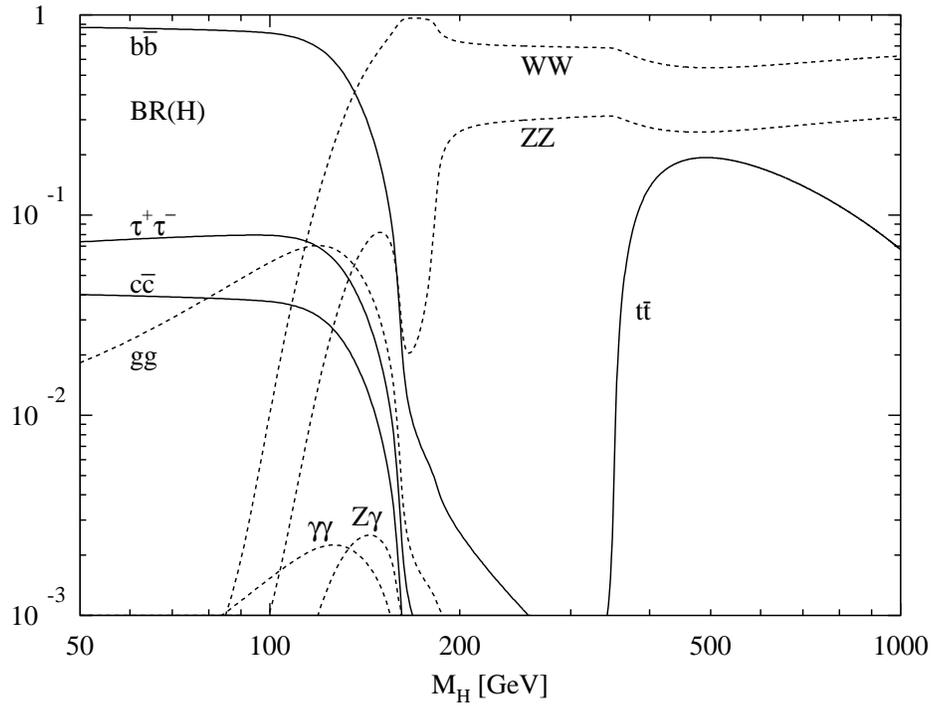}
\end{turn}
\vspace*{0.0cm}
\end{center}
\caption[]{\it Total decay width $\Gamma(H)$ in GeV and the main branching
ratios BR(H) of the SM Higgs decay channels.}
\end{figure}

\subsection*{2. Higgs searches at Present Colliders}

\subsubsection*{2.1 Searches at LEP}

The most comprehensive search of Higgs bosons done so far was undertaken 
by the LEP experiments. In the SM, the main production process is the 
so--called Bjorken or bremsstrahlung process \cite{R9}, where the $Z$ 
resonance emmits a Higgs boson, turns virtual and decays into two massless 
fermions
\beq
 \hspace*{1cm} Z \ra Z^* H \ra H f\bar{f} \nonumber
\eeq
Although the virtuality of the $Z$ boson is penalyzing since the cross
section is suppressed by a power of the electroweak coupling, the large
number of $Z$ bosons collected at LEP1 allows to have a sizeable rate
for not too heavy Higgs bosons. From the negative search of such events,
a lower bound of $M_H \gsim 65$ GeV has been set \cite{R2b}. Note that even
almost masseless Higgs bosons have been ruled out using this process:
indeed, even in this case, the Higgs particle will carry momentum and
will alter the kinematics of the visible final $Z^* \ra f\bar{f}$ state.  \s

At LEP2 \cite{R4}, 
with a center of mass energy above the $2M_W$ threshold, the
SM Higgs boson will be searched for using the previous process with
the difference that now the final $Z$ boson is on--shell. Although the 
production rates are much smaller than on the $Z$ resonance, the process 
is at lowest order in the electroweak coupling and gives a decent cross 
section. The Higgs bosons will mainly decay into $b\bar{b}$ final states, 
requiring efficient means to tag the $b$--quark jets. The backgrounds are 
rather small, except for the process $\ee \ra ZZ \ra b\bar{b}Z$ for Higgs 
masses close to $M_Z$. Depending on the final energy which will be reached 
at LEP2, $\sqrt{s}=175$ or $192$ GeV, Higgs masses close to 80 and 90 GeV, 
respectively, can be probed with an integrated luminosity of $\int {\cal L} 
=150$ pb$^{-1}$. The current Higgs mass bound  from these searches is $M_H 
\gsim 77$ GeV \cite{R2}. 

\subsubsection*{2.2 Searches at the Tevatron}

Currently, the Fermilab Tevatron collider \cite{R5} is operating at a c.m. 
energy $\sqrt{s}=1.8$ TeV with a luminosity ${\cal L} \sim 10^{31}$ 
cm$^{-2}$s$^{-1}$. With the main injector, which is expected to begin 
operation in a few years, the luminosity will be increase to ${\cal L} \sim 
2.10^{32}$ cm$^{-2}$s$^{-1}$ and the c.m. energy to $\sqrt{s}=2$ TeV. An 
increase of the luminosity to the level of ${\cal L} \sim 10^{33}$ 
cm$^{-2}$s$^{-1}$ [the so--called TEV33 option] is also currently 
discussed. \s

The most promising production mechanism of the SM Higgs boson at the 
Tevatron collider is the $W H$ process \cite{R10}, with the Higgs boson 
decaying into $b\bar{b}$ [or $\tau^+ \tau^-$] pairs
\beq
 \hspace*{1cm} qq \ra W H \ra W b \bar{b} \nonumber
\eeq
For Higgs masses $M_H \sim 100$ GeV, the cross section is of the order 
of a few tenths of a picobarn. The related production process $q \bar{q} 
\ra ZH \ra Z b\bar{b}$ [which is the equivalent of the Bjorken process 
in pp collisions] has a smaller cross section, a result of the small 
neutral current couplings compared to charged current  couplings. \s

The main irreeducible backgrounds will consist of $W b \bar{b}$ and $WZ 
\ra Wb\bar{b}$ for $M_H \sim M_Z$, as well as $t\bar{t}$ production for
$M_H \gsim 100$ GeV. These backgrounds have cross sections which are 
of the same order as the signal cross section; an important issue will be 
the $b\bar{b}$ invariant mass resolution which needs to be measured with 
a very good accuracy. The 
Higgs search at the Tevatron with a luminosity of $\sim 2$ fb$^{-1}$ will 
probably be limited to \cite{R5} $ M_H \lsim M_Z$, a mass region which will 
be already covered at LEP2. To probe Higgs masses larger than $M_Z$, a higher 
luminosity will be required, and the TEV33 option will be mandatory. A 
rather detailled analysis for TEV33 with an integrated luminosity  of 
$\int {\cal L} \sim 30$ fb$^{-1}$, concluded that Higgs masses up to $M_H 
\sim 120$ GeV could possibly be reached.  The processes $WH, ZH$ with 
$H \ra \tau^+ \tau^-$ and $W,Z \ra2$ jets will not significantly change
this picture. \s

\subsection*{3. Higgs Seraches at Future Colliders}

\subsubsection*{3.1 Production at LHC} 

The main production mechanisms of neutral Higgs bosons at hadron colliders
are the following processes:
\begin{eqnarray}
\begin{array}{lccl}
(a) & \ \ {\rm gluon-gluon~fusion} & \ \ gg  \ \ \ra & H \nonumber \\
(b) & \ \ WW/ZZ~{\rm fusion}       & \ \ VV \  \ra &  H \nonumber \\
(c) & \ \ {\rm association~with}~W/Z & \ \ q\bar{q} \ \ \ra & V + H \nonumber
\\
(d) & \ \ {\rm association~with~}\bar{t}t & gg,q\bar{q}\ra & t\bar{t}+H
\nonumber
\end{array}
\end{eqnarray}

In the interesting mass range, $100 \lsim M_H \lsim 250$ GeV, the dominant 
production process of the SM Higgs boson is the gluon--gluon fusion mechanism 
\cite{R11} [it is the case of the entire Higgs mass range] for which the cross 
section is of order a few tens of pb. 
It is followed by the $WW/ZZ$ fusion 
processes \cite{R11b} [especially for large $M_H$] 
with a cross section of a few pb; the cross sections of the associated
production with $W/Z$ \cite{R10} or $t\bar{t}$ \cite{R12} 
are an order of magnitude smaller. Note
that for a luminosity of ${\cal L}=10^{33} (10^{34})$~cm$^{-2}$s$^{-1}$, 
$\sigma=$~1 pb would correspond to $10^{4}(10^{5})$ events per year. \s

\begin{figure}[hbtp]
\begin{center}
\begin{turn}{-90}%
\epsfxsize=10.cm \epsfbox{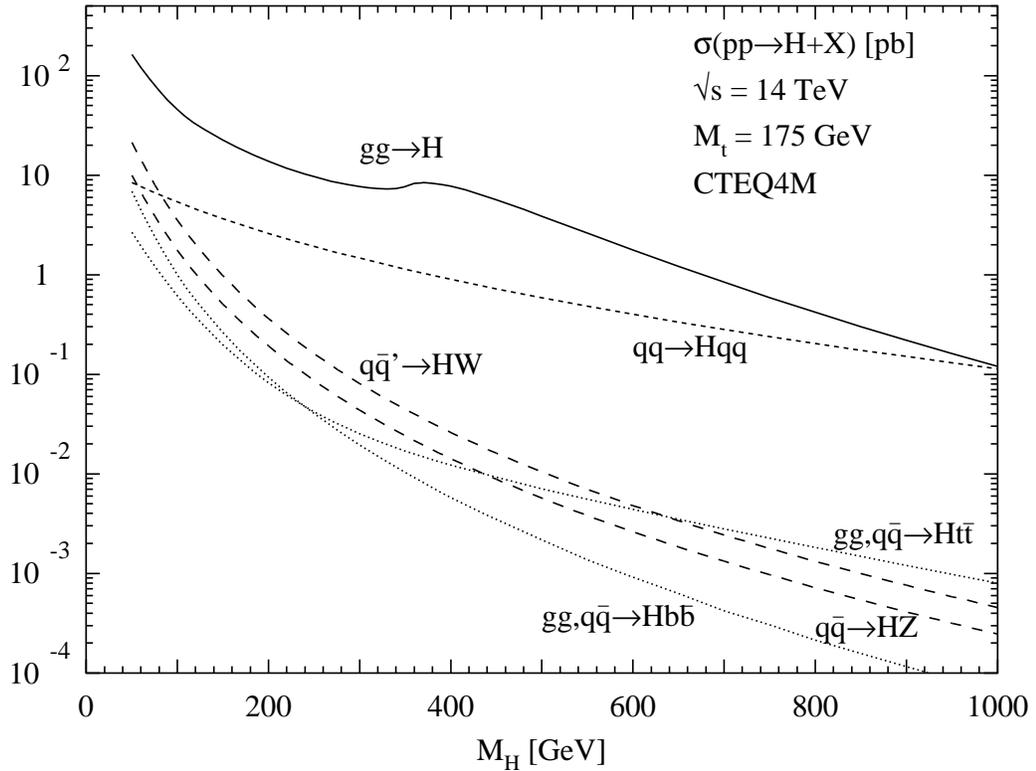}
\end{turn}
\vspace*{0.0cm}
\end{center}
\caption[]{\it  Production cross sections of the Higgs boson 
at LHC; from Ref.~\cite{Michael}.}
\end{figure}

Besides the errors due to the relatively poor knowledge of the gluon 
distribution at small $x$, the lowest order cross sections are affected 
by large uncertainties due to higher order corrections. Including the 
next to leading QCD corrections, the total cross sections can be defined 
properly: the scale at which one defines the strong coupling constant 
is fixed and the [generally non--negligible] corrections are taken into 
account. The ``K--factors" for $WH/ZH$ production [which can be inferred 
from the Drell--Yan $W/Z$ production] and the $VV$ fusion mechanisms 
are small, increasing the total cross sections by $\sim$ 20 and 10\% 
respectively \cite{R14} ;  the QCD corrections to the associated
$t\bar{t}H$ production are still not known. The [two--loop] QCD corrections 
to the main mechanism, $gg \ra H$, have been computed \cite{R13} and 
found to be rather large since they increase the cross sections by a 
factor $\simeq 1.8$ at LHC [there is, however, an uncertainty of $\sim 20\%$ 
due to the arbitrariness of the choice of the renormalization and 
factorization scales and also of the parton densities]. \s

The signals which are best suited to identify the produced Higgs particles
at the LHC have been studied in great detail in Ref.~\cite{R6}. I briefly
summarize here the main conclusions of these studies.\s

For Higgs bosons in the ``high mass" region, $M_H \gsim 130$~GeV, the signal
consists of the so--called ``gold--plated" events $H \ra Z Z^{(*)} \ra 4l^\pm$
with $l=e,\mu$. The backgrounds [mostly $pp \ra ZZ^{(*)}, Z \gamma^*$ for the 
irreducible background and $t \bar{t} \ra WWb \bar{b}$ and $Zb \bar{b}$ for the 
reducible one] are relatively small. One can probe Higgs masses up to ${\cal 
O}$(700~GeV) with a luminosity $\int {\cal L}= 100 $~fb$^{-1}$ at LHC. 
The $H \ra WW^{(*)}$ decay channel is more difficult to use because of the 
large background from $t\bar{t}$ pair production; the $H \ra t\bar{t}$
signal is swamped by the irreducible background from $gg\ra t\bar{t}$. 
For $M_H \gsim 700$ GeV
[where the Higgs boson total decay width becomes very large], the search 
strategies become more complicated; see Ref.~\cite{R7}. \s

For the ``low mass" range, the situation is more complicated. The branching
ratio for $H\ra ZZ^*$ becomes too small and due to the huge QCD jet background,
the dominant mode $H\ra b\bar{b}$ is practically 
useless; one has then to rely on the rare
$\gamma \gamma$ decay mode with a branching ratio of ${\cal O}(10^{-3})$. At
LHC with a luminosity of $\int {\cal L}= 100$~fb$^{-1}$, the cross section
times the branching ratio leads to ${\cal O}(10^{3})$ events but one has to
fight against formidable backgrounds. Jets faking photons need a rejection
factor larger than $10^{8}$ to be reduced to the level of the physical
background $q\bar{q}, gg \ra \gamma \gamma$ which is still very large. However,
if very good geometric resolution and stringent isolation criteria, combined
with excellent electromagnetic energy resolution to detect the narrow $\gamma
\gamma$ peak of the Higgs boson are available [one also needs a high
luminosity $ {\cal L} \simeq 10^{34}$~cm$^{-2}$s$^{-1}$], this channel,
although very difficult, is feasible: for $\int {\cal L}=100$ fb$^{-1}$,
ATLAS claims a sensitivity for $110 \lsim M_H \lsim 140$ GeV and requires
five times more luminosity to reach down masses $M_H \sim 80$ GeV; CMS
[which benefits from a good electromagnetic calorimeter] claims a 
coverage  $85 \lsim M_H \lsim 150$ GeV for 100 fb$^{-1}$. The low end
of the mass range is the most challenging due to the small branching ratio
and the larger backgrounds. \s

Complementary production channels would be the $pp \ra WH, t\bar{t}H 
\ra \gamma \gamma l \nu$ processes for which the backgrounds are much smaller
since one requires an additional lepton. However the signal cross sections 
are very small too making these processes also difficult. The processes $pp 
\rightarrow WH$ and $t\bar{t}H$ with $H \ra b\bar{b}$ 
seem also promising provided that very good micro--vertexing to tag the 
$b$--quarks can be achieved \cite{R15a}. Another complementary detection channel
in the intermediate mass range is $H \ra WW^*$ \cite{R15b}
which has a larger branching
ratio than  $H \ra ZZ^*$; the correlations among the final particles
would help to extract the signal from the $WW$ background. 
 
\subsubsection*{3.2 Production at e$^+$e$^-$ Colliders}

At $\ee$ linear colliders operating in the 500 GeV energy range  the
main production mechanisms for SM Higgs particles are \cite{R7}
\begin{eqnarray}
\begin{array}{lccl}
(a)  & \ \ {\rm bremsstrahlung \ process} & \ \ \ee & \ra (Z) \ra Z+H \non \\
(b)  & \ \ WW \ {\rm fusion \ process} & \ \ \ee & \ra \bar{\nu} \ \nu \
(WW) \ra \bar{\nu} \ \nu \ + H \non \\
(c)  & \ \ ZZ \ {\rm fusion \ process} & \ \ \ee & \ra e^+ e^- (ZZ) \ra
e^+ e^- + H \non \\
(d)  & \ \ {\rm radiation~off~tops} & \ \ \ee & \ra (\gamma,Z) \ra
t \bar{t}+H \non
\end{array}
\end{eqnarray}

The Higgs--strahlung \cite{R9} 
cross section scales as $1/s$ and therefore dominates 
at low energies while the $WW$ fusion mechanism \cite{R11b,R16}
 has a cross section 
which rises like $\log(s/M_H^2)$ and dominates at high energies. 
At $\sqrt{s} \sim 500$ GeV, the two processes have approximately 
the same cross sections for the interesting range 100 GeV $\lsim M_H 
\lsim$ 200 GeV. With an in integrated luminosity $\int {\cal L}
\sim 50$ fb$^{-1}$, approximately 2000 events per year can be collected
in each channel; a sample which is more than enough to discover the
Higgs boson and to study it in detail. The $ZZ$ fusion mechanism $(c)$
and the associated production with top quarks $(d)$ \cite{R17} 
have much smaller 
cross sections. But these processes will be very useful when it comes 
to study the Higgs properties as will be discussed later. \s

\begin{figure}[hbtp]
\begin{center}
\vspace*{-2.0cm}
\begin{turn}{-90}
\epsfxsize=10.cm \epsfbox{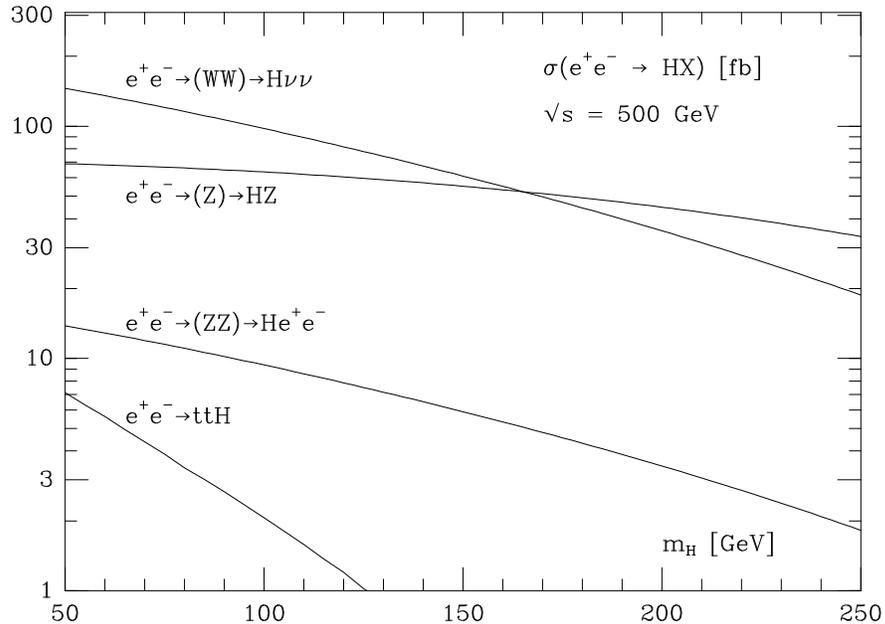}
\end{turn}
\vspace*{-0.5cm}
\end{center}
\caption[]{\it  Production cross sections of the Higgs boson 
in $e^+ e^-$ as a function of $M_H$.}
\end{figure}

In the Higgs-strahlung process, the recoiling $Z$ boson
[which can be tagged through its clean $\mu^+ \mu^-$ decay] is
mono-energetic and the Higgs mass can be derived from the energy of the
$Z$ if the initial $e^+$ and $e^-$ beam energies are sharp
[beamstrahlung, which smears out the c.m. energy should thus be suppressed as
strongly as possible, and this is already the case for machine designs
such as TESLA]. Therefore, it will be easy to separate the signal from
the backgrounds. For low Higgs masses, $M_H \lsim 130$ GeV, the
main background will be $\ee \ra ZZ$. The cross section is large, but it
can be reduced by cutting out the forward and backward directions 
[the process is
mediated by $t$--channel $e$ exchange] and by selecting $b\bar{b}$
final states by means of $\mu$--vertex detectors [while the Higgs decays
almost exclusively into $b\bar{b}$ in this mass range, BR$(Z \ra
b\bar{b}$) is small, $\sim 15\%$]. The background from single $Z$
production, $\ee \ra Zq\bar{q}$, is small and can be further reduced by
flavor tagging. In the mass range where the decay $H \ra WW^*$ is
dominant, the main background is triple gauge boson production and is
suppressed by two powers of the electroweak coupling. \s

The $WW$ fusion mechanism offers a complementary 
production channel. For small $M_H$, the main backgrounds are single $W$ 
production, $\ee \ra e^\pm W^\mp \nu$ $[W \ra q\bar{q}$ and the $e^\pm$ 
escape detection] and $WW$ fusion into a $Z$ boson, $\ee \ra \nu 
\bar{\nu}Z$, which have cross sections 60 and 3 times larger than the
signal, respectively. Cuts on the rapidity spread, the energy and
momentum distribution of the two jets in the final state [as well as 
flavor tagging for small $M_H$] will suppress these background events. \s

Additional processes are provided by the loop induced associated 
production of the Higgs boson with a photon \cite{R18} and 
double Higgs production \cite{R19}. The cross sections are very small 
and large luminosities will be required for these processes. \s

It has been shown in detailed simulations \cite{R7}, that just a few fb$^{-1}$ 
of integrated luminosity are needed to obtain a 5$\sigma$ signal for
a Higgs boson with a mass $M_H \lsim 140$ GeV at a 500 GeV collider 
[in fact, in this case, it is better to go to lower energies where the 
cross section is larger], even if it decays invisibly [as it could 
happen in SUSY models for instance]. Higgs bosons with masses 
up to $M_H \sim 350$ GeV can be discovered at the 5$\sigma$ level, in both 
the strahlung and fusion processes at an energy of 500 GeV and with a
luminosity of 50 fb$^{-1}$. For even higher masses, one needs to 
increase the c.m. energy of the collider, and as a rule of thumb, Higgs 
masses up to $\sim 70\%$ of the total energy of the collider can be 
probed. This means than a $\sim 1$ TeV collider will be needed to probe the 
entire Higgs mass range in the SM. 

\newpage

\subsection*{4. Study of Higgs properties}

Once the Higgs boson is found it will be of great importance to explore 
all its fundamental properties. This can be done at great details especially 
in the clean environment of $\ee$ linear colliders: the Higgs mass, the
spin and parity quantum numbers and the couplings to fermions and
gauge bosons can measured. In the following we will summarise these
feathures in the case of the SM Higgs boson.

\subsubsection*{4.1 Studies at $\ee$ Colliders}

In the Higgs--strahlung process with the $Z$ decaying into
visible particles, the mass resolution achieved with kinematical
constraints is close to 5 GeV, and a precision of about $\pm 200$ MeV
can be obtained on the Higgs mass with $\int {\cal L}=10$
fb$^{-1}$ if the effects of beamstrahlung are small \cite{C1}.
For masses below 250 GeV, the Higgs boson is extremely narrow and its
width cannot be resolved experimentally; only for higher masses [or at
$\mu^+ \mu^-$ colliders, see \cite{C2} e.g.] $\Gamma_H$ can be 
measured directly. \s

The angular distribution of the $Z/H$ in the Higgs--strahlung 
process is sensitive to the spin--zero of the Higgs particle: 
at high--energies the $Z$ is
longitudinally polarized and the distribution follows the $\sim
\sin^2\theta$ law which unambiguously characterizes the production of a
$J^P=0^+$ particle. The spin--parity quantum numbers of the Higgs
bosons can also be checked experimentally by looking at correlations in
the production $\ee \ra HZ \ra$ 4--fermions or decay $H \ra WW^* \ra$
4--fermion processes \cite{C3}, as well as in the more difficult
channel $H \ra \tau^+ \tau^-$ \cite{C4} 
for $M_H \lsim 140$ GeV. An unambiguous test
of the CP nature of the Higgs bosons can be made in the process $\ee \ra
tt \bar{H}$ \cite{C5} or at laser photon colliders in the loop 
induced process $\gamma \gamma \ra H$ \cite{C6}. \s

The masses of the fermions are generated through the Higgs
mechanism and the Higgs couplings to these particles are proportional to
their masses. This fundamental prediction has to be verified
experimentally. The Higgs couplings to $ZZ/WW$ bosons can be
directly determined by measuring the production cross sections in the
bremsstrahlung and the fusion processes. In the $\ee \ra H\mu^+\mu^-$
process, the total cross section can be measured with a precision of
less than 10\% with 50 fb$^{-1}$ \cite{C1}. \s

The Higgs couplings to light fermions are harder to measure,
except if $M_H \lsim 140$ GeV. The Higgs branching ratios to $b\bar{b}$,
$\tau^+\tau^-$ and $c\bar{c}+gg$ can be measured with a precision of
$\sim 5, 10$ and $40 \%$ respectively for $M_H \sim 110$ GeV \cite{C7}.
For $M_H \sim 140$ GeV, BR$(H \ra WW^*)$ becomes sizeable and can be
experimentally determined; in this case the absolute magnitude of the
$b$ coupling can be derived since the $HWW$ coupling is fixed by the
production cross section. 
The Higgs coupling to top quarks, which is the largest coupling in
the SM is directly accessible in the process $\ee \ra
t\bar{t}H$ \cite{R7}. For $M_H \lsim 130$ GeV, $\lambda_t$ can be
measured with a precision of about 10 to 20\% at $\sqrt{s}\sim 500$ GeV
with $\int {\cal L} \sim 50$ fb$^{-1}$. For $M_H \gsim 350$ GeV, the
$Ht \bar{t}$ coupling can be derived by measuring the $H \ra t\bar{t}$
branching ratio at higher energies \cite{C8}. \s

Finally, the measurement of the trilinear Higgs self--coupling,
which is the first non--trivial test of the Higgs potential, is
accessible in the double Higgs production processes $\ee \ra ZHH$ and
$\ee \ra \nu \bar{\nu}HH$ \cite{C9}. However, the cross sections are rather
small and very high luminosities [and very high energies in the second
process] are needed. 

\subsubsection*{4.2 Studies at the LHC}

In the ``low mass" range, the Higgs boson will appear as a very narrow
bump in the $\gamma \gamma$ invariant mass spectrum. The ATLAS and CMS 
collaborations \cite{R6} claim a $\gamma \gamma$ invariant mass resolution of 
approximately 1 GeV; so the Higgs boson mass will be measured with a
good accuracy if it is detected via its two--photon decay mode. \s

For masses above 250 GeV, the Higgs boson width will be greater than 
the experimental $4l^\pm$ resolution and can therefore be measured directly. 
This allows the determination of the $HWW$ and $HZZ$ couplings [assuming 
that are related by SU(2) custodial symmetry] since the $H \ra t\bar{t}$
branching ratio is rather small. Since the $4l^\pm$ rate is proportional
to $\sigma(gg \ra H) \times$ BR($H \ra ZZ)$, one could then determine
$\Gamma (H \ra gg)$ which allows to extract the $Ht\bar{t}$ coupling
[since the $Hgg$ couplings is dominantly mediated by the top quark loop 
contribution]. Some ratios of couplings could also be determined by
considering the processes $gg\ra H, qq \ra WH$ and $gg \ra t\bar{t}H$
with the subsequent decays $H \ra \gamma \gamma$ or $b\bar{b}$. 
 
\subsection*{5. Summary}

At the hadron collider LHC, the Standard Model Higgs boson can, in
principle, be discovered up to masses of ${\cal O}(1$~TeV). While the
region $M_H \gsim 130$ GeV can be easily probed through the $H \ra 4l^\pm$
channel, the $M_H \lsim 130$ GeV region is difficult to explore and a
dedicated detector as well as a high--luminosity is required to isolate
the $H\ra \gamma \gamma$ decay. \s

$\ee$ linear colliders with energies in the range of $\sim 500$ GeV are
ideal instruments to search for Higgs particles in the mass range below
$\sim 250$ GeV. The search for the Standard Model Higgs particle can be
carried out in several channels and the clean environment of the
colliders allows to investigate thoroughly its properties. 
Once the Higgs bosons are
found, the clean environment of $\ee$ colliders allows to study at
great details the fundamental properties of these particles. In this
respect, even if Higgs particles are found at LHC, high energy $\ee$ 
colliders will provide an important information which make them 
complementary to hadron machines.

\end{document}